\documentclass[journal]{IEEEtran}
\usepackage{hyperref}

\usepackage[utf8]{inputenc}
\usepackage{pgfplots} 
\usetikzlibrary{arrows, arrows.meta}
\pgfplotsset{compat=1.3}   
\usepackage{subcaption}
\ifCLASSINFOpdf
\else
\fi
\begin{document}
%
\title{Dual Frequency Comb Assisted Analog-to-Digital Conversion of Subcarrier Modulated Signals}
%
%
%

\author{Callum~Deakin,
        Zhixin Liu
\thanks{This work was supported by the EPSRC grant (EP/R041792/1), EPSRC programme grant TRANSNET (EP/R035342/1), Royal Society Paul Instrument Fund (PIF/R1/180001).}
\thanks{C. Deakin (email: callum.deakin.17@ucl.ac.uk) and Z. Liu (email: zhixin.liu@ucl.ac.uk) are with are with the Department of Electronic and Electrical
Engineering, University College London, London WC1E 7JE, UK.}
}

\markboth{\today}%
{Shell \MakeLowercase{\textit{et al.}}: Bare Demo of IEEEtran.cls for IEEE Journals}
%



\maketitle
\begin{abstract}
    Photonic analog to digital conversion offers promise to overcome the signal-to-noise ratio (SNR) and sample rate trade-off in conventional analog to digital converters (ADCs), critical for modern digital communications and signal analysis. We propose using phase-stable dual frequency combs with a fixed frequency spacing offset to downconvert spectral slices of a broadband signal and enable high resolution parallel digitization. To prove the concept of our proposed method, we demonstrate the detection of a 10-GHz subcarrier modulated (SCM) signal using 500-MHz bandwidth ADCs by optically converting the SCM signal to ten 1-GHz bandwidth signals that can be processed in parallel for full signal detection and reconstruction. Using sinusoidal wave based standard ADC testing, we demonstrate a spurious-free dynamic range (SFDR) of $>$45dB and signal-to-noise-and-distortion (SINAD) of $>$20dB, only limited by the receiver front-end design. Our experimental investigation reveals that this SINAD limitation can be overcome by improved receiver design, promising high resolution ADC for broadband signals.
\end{abstract}

\begin{figure*}[ht!]
\centering
\includegraphics[scale=0.5]{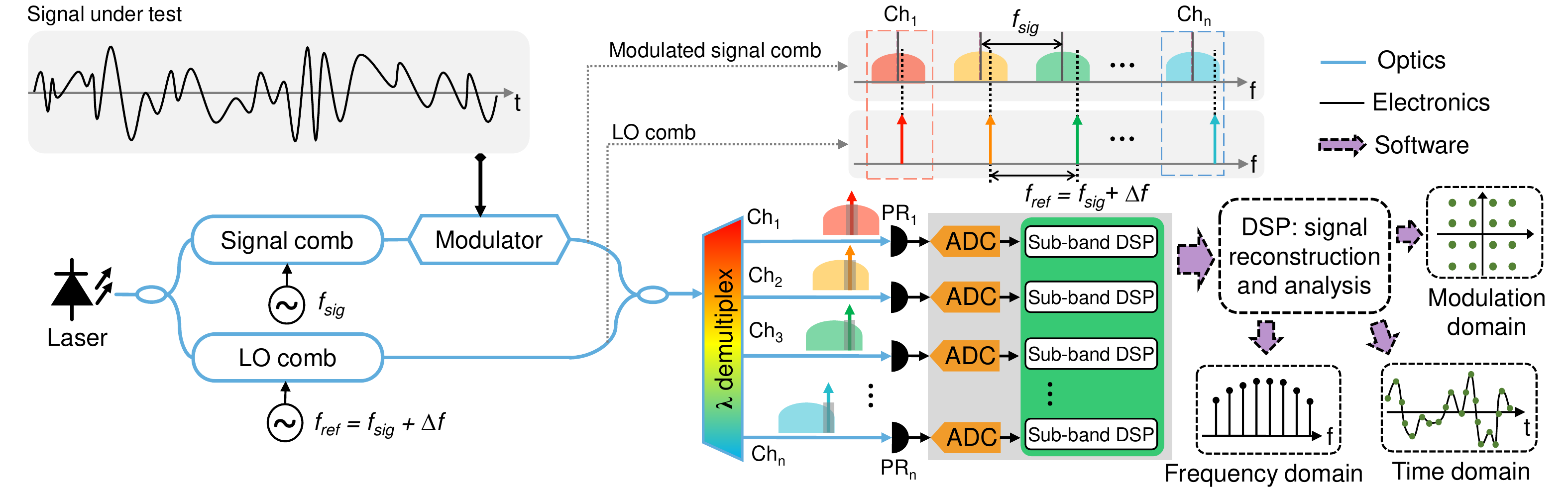}
\caption{Our envisaged system for the dual frequency comb assisted ADC. Beating between two frequency combs of spacing $f_{sig}$ and $f_{ref}$ can downconvert spectral slices of the signal under test to baseband with little distortion. These slices can then be detected at high resolution by a bank of parallel low speed ADCs. ADC: analog to digital converter; PR: photoreceiver; Ch:  channel; DSP: digital signal processing.}
\label{fig1}
\end{figure*}

Analog to digital converters (ADCs) are essential to the acquisition and processing of time varying electrical signals, being the basis of real-time oscilloscopes and enabling signal analysis across a broad spectrum of scientific and engineering disciplines. In particular, the exponential growth of optical and wireless communications over the past few decades has driven demand for high speed and high resolution ADCs at greater than 10~GHz bandwidth~\cite{keysight_infiniium_nodate,noauthor_teledyne_nodate}. 

High speed ADCs followed by tailored digital signal processing (DSP) is the typical approach when attempting high resolution detection of broadband signals. State-of-the-art ADCs can achieve up to 7 bits effective number of bits (ENOB), which corresponds to about 44~dB signal-to-noise ratio (SNR) at 10~GHz but this decays to around 5 bits when large bandwidths of over 70~GHz are required~\cite{keysight_infiniium_nodate-1}. The cause of this loss in resolution is threefold. Firstly, clock jitter (or phase noise) of the ADC's sampling clock increases quadractically with sampling frequency and therefore causes inevitable degradation of SNR at high sampling rates~\cite{rubiola_phase_2008}. Secondly, present semiconductor manufacturing limits transistor size (which is inversely proportional to speed) to around 5~nm with doubts as to whether die shrinks beyond 3~nm are viable~\cite{noauthor_international_nodate}. Thirdly, the fundamental `sample and hold' ADC method unavoidably results in frequency fading (thus reduced SNR) at high frequencies~\cite{clara2012high}.

Furthermore, the utilization of advanced DSP such as deep learning algorithms for the processing of ADC captured signals is reliant on small parallelization overheard as processors approach their upper bounds of single threaded instructions per second due to thermal and quantum effects~\cite{markov_limits_2014}. Thus, efficient serial to parallel conversion and data processing is required for for analysis of the captured data.

These challenges of high speed, high resolution signal detection as well as serial to parallel signal conversion have led to a variety of solutions across industry and academia. Leading instrument manufacturers have developed electronic time interleaving and digital/analogue bandwidth interleaving techniques to increase sampling rate and bandwidth~\cite{drenski2018adc}. More exotic solutions in academia include optical time stretching to improve ADC resolution and bandwidth~\cite{fischer_high-speed_2011,khilo_photonic_2012}, and optical Fourier transforms to enhance the digital signal processing capability~\cite{hillerkuss_simple_2010,hu_comb-assisted_2017}.

Many of these methods attempt to down convert spectral slices of a high bandwidth signal so that they can be processed by low speed high resolution ADCs in parallel. Dual frequency combs are ideal for this purpose, being widely used in spectroscopy to down convert the optical response of a material to the low frequency RF domain~\cite{coddington_dual-comb_2016}, and have been further utilized for low-noise, low distortion spectral replication~\cite{ataie_subnoise_2015,esman_detection_2016}. 

In this paper, we propose using the dual frequency comb approach to down convert high bandwidth signals for high resolution parallel digitization. As a proof of concept, we optically downconvert a 10-GHz bandwidth subcarrier modulated (SCM) signal, with each channel modulated with Nyquist-shaped 4-level pulse amplitude modulation (PAM4) format, to ten 1-GHz bandwidth SCM-PAM4 signals using two frequency combs with stable phase and a fixed frequency offset of 1GHz. The down-converted sub-band signals are captured using two 2.4-GSa/s high-resolution (14 bits) ADCs for balanced-detection of 500MHz baseband bandwidth. We achieve demodulation of the SCM-PAM4 signal of up to 20~dB SNR, limited primarily by the ENOB of the digital to analog converter (DAC) used. Using standard ADC performance testing method, we obtained a spurious-free dynamic range (SFDR) of 45dB across and a signal-to-noise-and-distortion (SINAD) of around 20dB, assuming 500MHz baseband bandwidth. To the best of our knowledge, this is the first demonstration of modulated signal detection using the dual comb method. Importantly, our work shows that the achievable SINAD is mainly limited by receiver front-end, which could be further improved for potentially significant improvement of the system performance.

The schematic diagram of our proposed photonic ADC scheme is shown in Fig.~\ref{fig1}. Two frequency combs with tone spacing of $f_{sig}$ and $f_{ref}$ are generated from the same seed laser source, which ensures that the combs are mutually coherent. The `signal' comb ($f_{sig}$) is modulated by driving a Mach-Zehnder modulator (MZM) with the signal under test and recombined with the `local oscillator (LO)' ($f_{ref} = f_{sig} + \Delta f$) comb before being demultiplexed by an arrayed waveguide grating (or other means of optical filtering) with a channel spacing of $f_{sig}$.

When the $n$-th channel is incident on a photodiode, the beating between the $n$-th tone of the LO comb and the modulated $n$-th tone of the signal comb will generate a baseband signal centered at a frequency of $n\Delta f$. With an appropriate anti-aliasing filter, a bank of $n$ ADCs can therefore be used to interrogate $n$ $\Delta f$ spectral slices of the original signal under test. Importantly, the optical phase of the two frequency combs needs to be stable over the entire capturing period to facilitate waveform recovery. 

There are several benefits to this approach. Firstly, the parallel detection of the sub-bands allows for digitizing high frequency components using low-speed but high-resolution ADCs. The whole signal spectrum can therefore be reconstructed at a resolution beyond the fundamental limit of a single wide-bandwidth ADC. Secondly, the optical spectral decomposition allows for parallel DSP of the sub-bands without any overhead for serial to parallel conversion, enabling reduced latency in the DSP chain and potential simplification of DSP hardware implementation. Thirdly, compared to digital/analog interleaving and time domain stretching techniques, our technique is not subject to the limitations imposed by sampling clock jitter since the bank of low speed ADCs are referenced from the same low frequency, low phase noise clock source. Finally, this approach is scalable: digitizing a signal of bandwidth $B$ via $N$ sub-bands requires $N$ comb lines with $\Delta f > B/N$, provided that $f_{sig}/2 > B$. Thus a high bandwidth signal can be processed simply by increasing the comb spacing $f_{sig}$ and by either increasing the number of channels $N$ or the sub channel bandwidth $\Delta f$. The only limitation to bandwidth therefore is being able to generate sufficient comb lines at a wide enough spacing to accommodate the signal under test. Considering recent demonstrations of ultra-wide band frequency combs with free spectral ranges of well over 100~GHz~\cite{foster_silicon-based_2011,okawachi_octave-spanning_2011}, and in some cases up to 850~GHz~\cite{delhaye_octave_2011}, our proposed photonic ADC approach has the potential to scale to several hundred GHz bandwidth.

\begin{figure*}[ht!]
\centering
\includegraphics[scale=0.55]{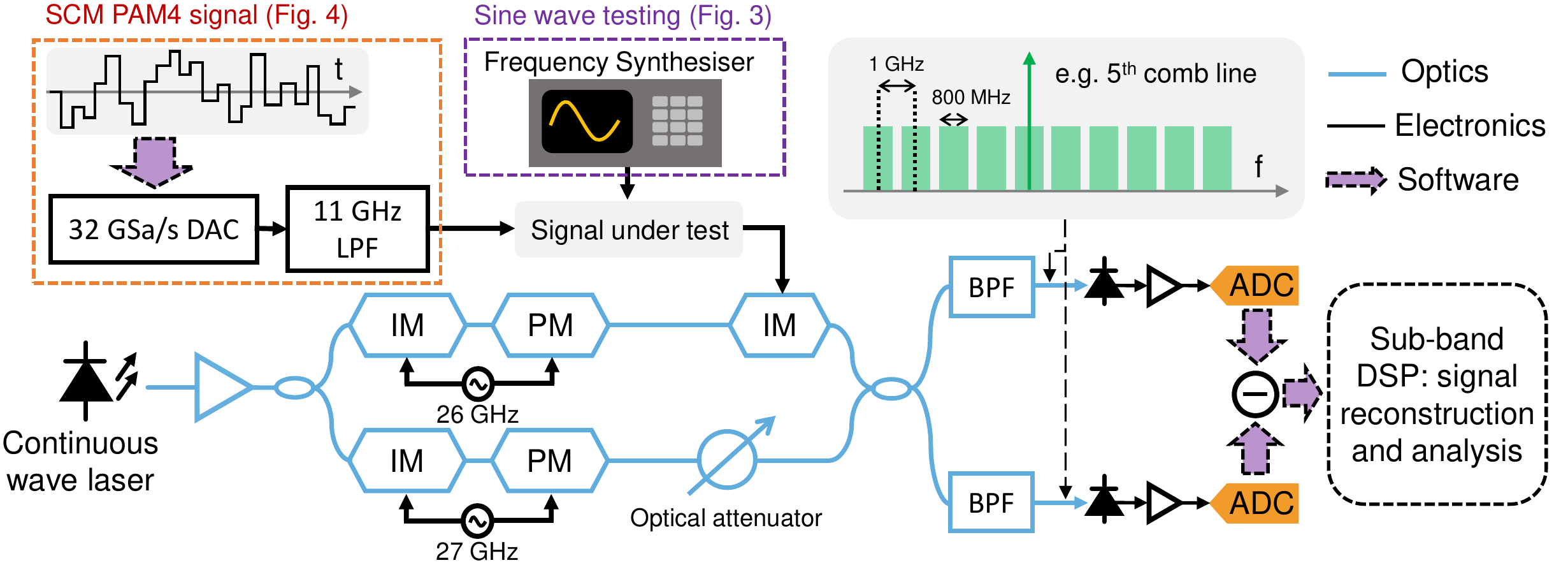}
\caption{Experimental setup for the results presented in this paper. The sub-bands are detected sequentially by a pair of 14-bit 2.4~GSa/s ADCs arranged as a balanced detector. Two 26~GHz tunable bandpass filters are used to select desired 1~GHz sub-band. BPF: bandpass filter, LPF: low pass filter, DAC: digital to analog converter, IM: intensity modulator, PM: phase modulator.}
\label{fig2}
\end{figure*}

The experimental setup is shown in Fig.~\ref{fig2}. In our proof-of-concept experiment, we generate both the signal comb and LO combs by driving cascaded phase and intensity modulators with sinusoidal RF signals at 26~GHz and 27~GHz, respectively, generating 24 tones with $<$~3dB power variation as in Ref.~\cite{torres-company_lossless_2008}. The combs were seeded with a narrow linewidth (5kHz) continuous wave (CW) signal amplified to 3W, resulting in an average power of 8 dBm per tone and $>$55 dB optical signal-to-noise ratio (OSNR). The signal comb was modulated by signal under test via an MZM biased at transmission null, which linearly mapped the electronic signals to all modulated tones to the optical domain with suppressed optical carriers. The peak-to-peak driving voltage was kept less than $0.3V\pi$ to ensure a highly linear mapping from the electronic to the optical domain. The modulated signal comb has about -15dBm per channel and was combined with the LO using a 50/50 coupler and subsequently filtered by two 30-GHz bandwidth tunable bandpass filters (BPFs) before detected by two 1.2-GHz photodiodes (PDs) followed by transimpedance amplifiers (TIA). By tuning the center wavelength of the BPFs, different sub-bands can be selected. The LO power was attenuated by 12dB to avoid the saturation of the TIA used in this experiment. This attenuation can be avoided for a better SNR performance in practical systems, as we will discuss later.  

The detected signals were captured by a pair of 14-bit 2.4~GSa/s ADCs arranged as a balanced detector to sequentially detect the 1~GHz spectral slices by tuning the filters to the desired sub band. Full detection of each sub-band would require coherent detection, but this approach is sufficient to detect a series of sub-band limited amplitude modulated signals as we demonstrate in this paper. The fiber lengths of the two arms (signal comb and LO comb) were matched in our system. The fibers in both signal comb and LO comb branches are attached to each other to minimize the impact from environment, ensuring optical phase only slowly drifted with temperature, allowing for the detection of the combined signals with stable optical phase over the entire period (2.048 $\mu$s) of the SCM-PAM4 signals. 

 As shown in the insets in Fig.~\ref{fig2}, the signal under test was generated by using either a 6-bit 32-GSa/s digital-to-analog converter (DAC) for the generation of the subcarrier modulated (SCM) signals, or a frequency synthesizer to generate a range of single frequency sine wave for SFDR and SINAD testing based on the IEEE ADC testing standard~\cite{5692956}. In the DAC setup, a 11~GHz low pass filter was used to remove the aliases. The RF driving power was kept at 0~dBm for optimum performance. The SCM signal under test was a 10 channel, 1~GHz spaced, 800~MBaud sub carrier modulated Nyquist-shaped PAM-4 signal. The SCM modulation format allows for independent detection and decoding of each subchannel without requiring parallel detection of the subbands. A 40 MHz frequency shift was added to the baseband of each channel to avoid the low frequency filtering (DC-10 MHz) caused by the AC-coupled post detection electrical amplifier used in our experiment. This allows for an assessment of the system performance without the significant investment of acquiring sufficient ADCs to detect a signal with a full bandwidth baudrate. The signal demodulation and analysis was digitally processed offline, using a two sample-per-symbol 17-tap feed-forward equalizer to equalize the frequency roll-off within the sub-band. Finally, for performance evaluation, we also captured the SCM-PAM4 directly from the DAC output using a Tektronix DPO72304DX digital phosphor oscilloscope with a sampling rate of 100~GSa/s and a 3~dB analog bandwidth of 23~GHz.

\begin{figure}[htb]
\centering
\begin{subfigure}[b]{\linewidth}
\centering
\begin{tikzpicture}[trim axis left,trim axis right]
\begin{axis} [ylabel=Power ratio (dB), xlabel=Frequency (GHz),
ymin=0,ymajorgrids=false,xmajorgrids=true,legend pos=south east,xmin=0,
axis y line*=left,,height=0.5\linewidth,width=0.9\linewidth,legend columns=2]
\addplot table [x=Freq(GHz), y=SFDR, col sep=comma] {SFDRs.csv};
\addlegendentry{SFDR}
\addplot table [x=Freq(GHz), y=SINAD, col sep=comma] {SFDRs.csv};
\addlegendentry{SINAD}
\node at (axis cs:0.5, 5) {(a)};
\end{axis}
\begin{axis}[
        ymin=0,
        ymax=55,
        axis x line=none,
        axis y line*=right,
        ylabel=ENOB (bits),
        y coord trafo/.code=\pgfmathparse{(#1-1.76)/6.02},
        height=0.5\linewidth,width=0.9\linewidth
    ]
    \addplot [draw=none,forget plot] table [x=Freq(GHz), y=ENOB, col sep=comma] {SFDRs.csv};
    \end{axis}
\end{tikzpicture}

\end{subfigure}%
\\
\begin{subfigure}[b]{\linewidth}
\centering
\begin{tikzpicture}[trim axis left,trim axis right]
\begin{axis} [xmax=0.5,xmin=0,ymax=10,ymin=-70,ylabel=Power (dBc), xlabel=Frequency (GHz),,height=0.5\linewidth,width=0.9\linewidth]
\addplot table [x=freqs, y=PSD, col sep=comma,mark=none] {exampleSine.csv};
\draw[|-|] (axis cs:0.21, 0) -- (axis cs:0.21, -44.11);
\node at (axis cs:0.17, -20) {SFDR};
\draw[|-|] (axis cs:0.4, 0) -- (axis cs:0.4, -20.7521);
\node at (axis cs:0.35, -10) {SINAD};
\draw[|-|] (axis cs:0.42, -20.7521) -- (axis cs:0.42, -59.88);
\node at (axis cs:0.34, -40) {Process gain};
\node at (axis cs:0.1, 0) {RBW = 61~kHz};
\node at (axis cs:0.02, -45) {(b)};
\end{axis}
\end{tikzpicture}
\end{subfigure}
\caption{(a) SFDR and SINAD/ENOB as a function of frequency up to 10.5~GHz, with (b) an example fast fourier transform (FFT) showing a 5.25~GHz signal detected in the 4.5-5.5~GHz sub-band. All values are calculated over a 1~GHz range after digital Nyquist filtering with an average of 4 16384-point FFTs, corresponding to an FFT process gain of 39.1~dB and resolution bandwidth (RBW) of 61~kHz.}\label{fig3}
\end{figure}
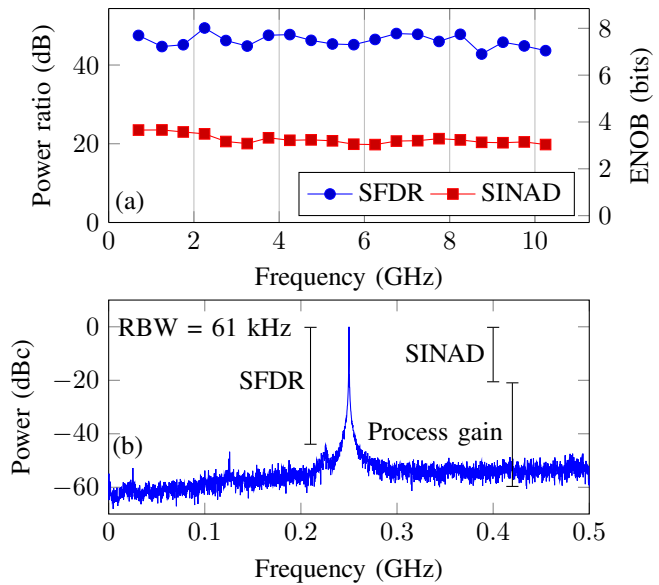

We first investigate the system performance using sinusoidal waves based on the IEEE standard. Fig.~\ref{fig3}(a) shows the SFDR and the SINAD, as well as the effective number of bits, $\textnormal{ENOB} = (\textnormal{SINAD} - 1.76)/6.02$. The SFDR remains consistently above 45~dB and the SINAD above 20~dB for almost the entire bandwidth, with both figures of merit experiencing a gradual roll off of around 3~dB across the measurement bandwidth. This decrease in performance with higher frequency can mainly be attributed to the high frequency-roll off from 0.1 to 10~GHz in the electrical signal chain (inc. RF cables and connectors), which was about 3~dB in our system. Since the optical bandwidth of each subchannel is 1 GHz, only 500~MHz electronic base band bandwidth is needed for the signal detection. Thus, the SINAD was calculated by integrating the noise and distortion power from 0-500 MHz.

Fig. \ref{fig3}b shows the RF spectrum calculated from the captured subcarrier signals when the signal under test was a 5.25~GHz sinusoidal signal. The BPFs were tuned to select the 5th subchannel (4.5-5.5~GHz), which is displayed as an average of four 16384-point FFTs corresponding to an FFT process gain of 39.1~dB. Note that because we do not perform full coherent detection of the sub-band, the two halves of each 1~GHz sub-band are aliased into the same 500~MHz half sub-band. Thus a 4.75~GHz signal will appear as the same 250~MHz sub-band signal as the 5.25~Ghz signal shown in Fig.~\ref{fig3}(b).

The limitation of our current system, as for other dual comb based signal processing methods with a similar concept, can be illustrated by analyzing Fig.~\ref{fig3}(b). The SFDR is limited by the presence of spurs caused by receiver-side electrical amplifiers. Furthermore, the used PD-TIA saturated at a low input optical power of -13~dBm, which prevented the full use of the ADCs dynamic range and a significant drop in the measured SINAD. This is highlighted by the large disparity between the measured SFDR and SINAD. The above-mentioned engineering problems can be resolved with the optimization of the receiver front-end design, such as the transimpedance and the saturation power of the TIA as well as the 2nd stage voltage amplifier gain to obtain a potentially significant improvement (e.g. a minimum of 12dB from higher saturation power) in the SINAD and ENOB. 

\begin{figure}[htb]
\centering
\begin{subfigure}[b]{\linewidth}
\centering
\begin{tikzpicture}
\begin{axis} [ylabel=SNR (dB), xlabel=Channel no.,xtick={0,1,2,3,4,5,6,7,8,9,10},ymin=10,ymajorgrids=false,xmajorgrids=false,height=0.6\linewidth,width=\linewidth,legend pos=south east]
\addplot table [x=Channel, y=SNR, col sep=comma] {PAM4_SNRs.csv};
\addlegendentry{Photonic ADC}
\addplot table [x=Channel, y=TekSNR, col sep=comma] {PAM4_SNRs.csv};
\addlegendentry{Tektronix}
\node at (axis cs:0.5, 22) {(a)};
\draw (axis cs: 5, 17.8) circle [radius=0.2cm] ;
\draw (axis cs:4.05, 17.6) -- (axis cs:4.7, 17.8);
\end{axis}
 \begin{axis}[xshift=.03\textwidth,yshift=0.05\textwidth,width=0.45\textwidth,yticklabels={,,},xtick={-1,1},xmin=-1.4,xmax=1.4]
  \addplot table [x=time, y=amp, col sep=comma,,mark=none] {histplot.csv};
 \end{axis}
\end{tikzpicture}

\end{subfigure}
\begin{subfigure}[b]{\linewidth}
\centering
\begin{tikzpicture}
\begin{axis} [ylabel=Power (dBc), xlabel=Frequency (GHz),xmin=0,xmax=1.2,ymin=-40,ymax=15,height=0.5\linewidth,width=\linewidth]
\addplot table [x=freqs, y=PSD, col sep=comma,mark=none] {detectedSpectrum.csv};
\draw[dashed](axis cs:0.5, -40) -- (axis cs:0.5, 15);
\draw[>=triangle 45,<->](axis cs:0,3) -- (axis cs:0.5, 3);
\node at (axis cs:0.25, 9) {Detected sub-band};
\draw[>=triangle 45,<->](axis cs:0.5,3) -- (axis cs:1.2, 3);
\node at (axis cs:0.85, 8.5) {Filtered adjacent channels};
\node at (axis cs:0.1, -35) {(b)};
\end{axis}
\end{tikzpicture}
\end{subfigure}
\caption{10 channel SCM PAM4 detection performance with 1~GHz sub-bands. (a) SNR vs channel number for our photonic ADC and a Tektronix DPO72304DX 100~GSa/s 23~GHz oscilloscope, with example sample histogram inset from the 5th channel. (b) Example sub-band spectrum from the 5th sub-band showing the target 5th channel and adjacent channels across the full ADC bandwidth.}\label{fig4}
\end{figure}
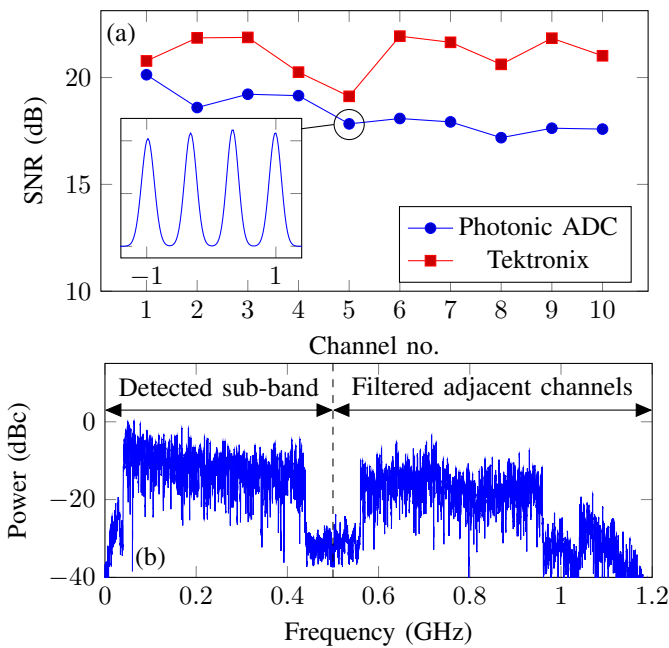

The measured SNR values for the 10-GHz bandwidth SCM-PAM4 signal are shown in Fig.~\ref{fig4}(a), with an example sampling histogram inset from which the SNR values are derived. The highest SNR we obtained was 20.1~dB in the first channel (0.5-1.5~GHz). An increase in channel number shows a decrease in SNR, with the final 10th channel (at 10~GHz) suffering a 2.5~dB SNR penalty with reference to the first channel. This SNR trend agrees with that demonstrated for SINAD/ENOB in Fig.~\ref{fig3}(a), with a further penalty of approximately 3~dB to the absolute SNR due to inter-channel interference, since transmitting a single sub-band from the arbitrary waveform generator allows us to recover the higher SNR/SINAD demonstrated in Fig.~\ref{fig3}(a). In addition, the achievable SNR in this experiment was fundamentally limited by the ENOB of the arbitrary waveform generator used to produce the SCM signal, which is highlighted by the comparison with the conventional Tektronix DPO72304DX oscilloscope. The maximum achievable SNR with direct detection by the oscilloscope is around 22~dB (3.4 ENOB) which is in line with our previous experience of the DAC perfomance. Noticeably, the capture with the oscilloscope does not experience any significant SNR roll off as the oscilloscope noise floor is 10-dB below that of the DAC. 

Fig.~\ref{fig4}(b) shows the spectrum of the received sub-band signal. Visible is the target 5th channel as well as the adjacent channels that are filtered out digitally, with further channels are suppressed by the ADC anti-aliasing filter who effects can be seen above 1~GHz. The strong (8~dB) roll off of the channel of interest is a result of the frequency response of the detector and is corrected in this experiment by the previously described equalizer.  

In summary, we propose an analog to digital conversion method based on phase-stable dual frequency combs. Our system down converts a sub-band of a broadband signal optically for detection with high-resolution ADCs, promising potential to overcome the resolution and bandwidth tradeoff in the conventional time-interleaving ADC paradigm. The feasibility of the proposed method was experimentally demonstrated by detection of a 10-GHz SCM-PAM4 signal with 500-MHz bandwidth ADCs, showing 20dB SNR that is only limited by the signal generator in our experiment. We identify that the moderate SFDR and SINAD performance in our current system is caused by the receiver front-end and highlight routes for future high resolution photonic-assisted ADCs. 

\ifCLASSOPTIONcaptionsoff
  \newpage
\fi

\bibliographystyle{ieeetr}
\bibliography{IEEEabrv,paper}

\end{document}